\documentclass[twocolumn,showpacs,prl]{revtex4}
\usepackage{graphicx}
\begin{document}

\title{Effects of nonintegrability on stabilization of Feshbach
molecules in atom waveguides}

\author{V. A. Yurovsky}

\affiliation{School of Chemistry, Tel Aviv University, 69978 Tel Aviv,
Israel}

\date{\today}

\begin{abstract}Deactivation of broad quasi-one-dimensional dibosonic
molecules is analyzed. Within integrable Lieb-Liniger-McGuire (LLMG)
 model
an exact expression does not demonstrate suppression of the
 deactivation at
low collision energies. Solution of Faddeev equations demonstrates
 that
when a Feshbach resonance lifts the symmetry of the LLMG model the
deactivation becomes suppressed. This effect shows a way for
 formation of a
stable gas of dibosonic Feshbach molecules.\end{abstract}

\maketitle

Ultracold molecules have recently been formed by Feshbach resonance
association (see reviews \cite{TTHK99} and original works
\cite{Bose_Fesh,Fermi_Fesh}). These Feshbach molecules are very broad
 since
they are superpositions of the closed and open channels. However,
 inelastic
collisions of atoms and molecules lead to strong losses of molecules
composed from Bose atoms in experiments \cite{Bose_Fesh}. In the
 fermionic
case \cite{Fermi_Fesh} inelastic collisions are strongly suppressed
 due to
Pauli blocking and weak coupling of the atoms \cite{PSS}. This effect
allows investigation of BEC-BCS crossover and other phenomena of
fundamental physical importance.

Tight confinement of atomic motion in two directions by atom
waveguides strongly modifies atomic collisions in
 quasi-one-dimensional
(1D) regime, when the transverse excitation energy $\omega _{\perp }$
  (in units where $\hbar =1$)
substantially exceeds the collision energies \cite{O98}. This regime
 has
been realized in 2D optical lattices \cite{OptLatt,Mol1D}, elongated
 atomic
traps \cite{LongTrap}, and atomic integrated optics devices
\cite{AtIntOpt}. Broad quasi-1D molecules, predicted in \cite{BMO03}
 and
observed in \cite{Mol1D}, have binding energies less then $\omega
 _{\perp }$. Such diatoms
can be described by two-channel 1D model \cite{Y05,Y06}, where the
 closed
channel incorporates both 3D closed channel and excited transverse
waveguide modes.

Rates of deactivation into tightly bound (non-Feshbach) states are
approximately proportional to the probability to find three atoms
(two of
which belong to the molecule) in the same place, e. g. to three-body
(3B)
correlations \cite{PSS}. 3B correlations were analyzed in
 \cite{Gangardt_03}
within exactly-soluble Lieb-Liniger model \cite{LL63}. In this model
 1D
bosons interact by zero-range potentials $U_{a}\delta \left(
 z_{j}-z_{l}\right) $, where $z_{j}$  are atomic
coordinates and the interaction strength $U_{a}$  is
 energy-independent. In the
case of repulsive interactions ($U_{a}>0$), which does not bound the
 atoms, the
correlations are suppressed at low collision energies and strong
 interactions
\cite{Gangardt_03}. Bound states can be formed in the case of
 attractive
interactions ($U_{a}<0$) described by the McGuire solution
 \cite{McGuire64}. The
tree-atom correlations for atom and diatom with relative momentum
 $p_{0}$  can be
represented, using 3B wavefunction $\varphi _{0}\left(
 z_{1},z_{2},z_{3}\right) $ \cite{McGuire64}, as
\begin{equation}
|\varphi _{0}\left( 0,0,0\right) |^{2}={m|U_{a}|\over 24\pi { }
 ^{2}}{9p^{2}_{0}+m^{2}U{ } ^{2}_{a}\over p^{2}_{0}+m^{2}U{ }
 ^{2}_{a}} , \label{phiLLMG}
\end{equation}
where $m$ is the atomic mass. This expression has a non-zero limit at
low collision energies ($p_{0}\rightarrow \infty $) or strong
 interactions ($|U_{a}|\rightarrow \infty $). This
behavior differs from the free-atom case \cite{Gangardt_03}, since
 bound
atoms keep non-vanishing imaginary momenta $\pm {i\over 2}mU_{a}$
 even when $p_{0}\rightarrow 0$. Thus,
inelastic collisions of 1D broad molecules are cardinally different
 from
collisions of two structureless particles, which are suppressed both
within the Lieb-Liniger-McGuire (LLMG) model \cite{Gangardt_03} and
beyond it \cite{YB07}. This difference reflects the fact that broad
Feshbach molecules can not be considered as zero-range objects and
 their
deactivation is not a 2B process.

The present Letter demonstrates that deactivation of broad 1D
molecules becomes suppressed when integrability of LLMG model is
 lifted by
a Feshbach resonance. Similar effects can be expected for other
 mechanisms
of integrability lifting, e. g. due to virtual transverse mode
 excitation
\cite{MEG05,SCKB05}. In contrast with these effects, other processes,
 such
as reflection and dissociation in atom-diatom collisions and
 three-atom
association \cite{YBO06}, become allowed when the nonintegrability is
lifted.

Except of demonstration of a new observable effect of
 non-integrability,
which is interesting for atomic, molecular, and statistical physics,
 as well
as for quantum field theory, present results show a way for formation
 of a
stable gas of dibosonic Feshbach molecules.

\begin{figure}
\includegraphics[width=2.5in]{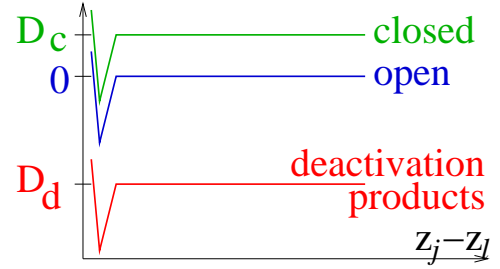}

\caption{Schematic description of 2B channel potentials.}
\label{Fig_chan}

\end{figure}

Consider multichannel collisions of 1D Bose atoms described by the
annihilation operators $\hat{\Psi }_{a}\left( z\right) $. The model
 includes several 2B channels (see
Fig.\ \ref{Fig_chan}). The Feshbach closed-channel state, described
 by the
molecular annihilation operator $\hat{\Psi }_{c}\left( z\right) $,
 lies at the energy $D_{c}$  close to the
open channel threshold, which serves as the energy origin. A set of
deactivation product channels $\{d\}$, described by the molecular
 annihilation
operators $\hat{\Psi }_{d}\left( z\right) $, lie at the energies
 $D_{d}$  far below the open channel
threshold. The system can be described by the Hamiltonian
\begin{eqnarray}
\hat{H}=\int dz\biggl\{\hat{\Psi }^{\dag }_{a}\left( z\right)
 \left\lbrack -{1\over 2m} {\partial { } ^{2}\over \partial z{ }
 ^{2}}+{U{ } _{a}\over 2}\hat{\Psi }^{\dag }_{a}\left( z\right)
 \hat{\Psi }_{a}\left( z\right) \right\rbrack \hat{\Psi }_{a}\left(
 z\right)  \nonumber
\\
+\sum\limits^{}_{j=c,\{d\}}\hat{\Psi }^{\dag }_{j}\left( z\right)
 \left( -{1\over 4m}{\partial { } ^{2}\over \partial z{ } ^{2}}
+D_{j}\right) \hat{\Psi }_{j}\left( z\right)
+\hat{V}_{\text{Fesh}}\left( z\right)  \nonumber
\\
+\hat{V}^{\dag }_{\text{Fesh}}\left( z\right)
+\sum\limits^{}_{\{d\}}\left\lbrack \hat{V}_{ad}\left( z\right)
+\hat{V}^{\dag }_{ad}\left( z\right) +\hat{V}_{cd}\left( z\right)
+\hat{V}^{\dag }_{cd}\left( z\right) \right\rbrack \biggr\} .
 \label{Hacd}
\end{eqnarray}
Here the interaction $\hat{V}_{\text{Fesh}}\left( z\right)
 =g\hat{\Psi }^{\dag }_{c}\left( z\right) \hat{\Psi }_{a}\left(
 z\right) \hat{\Psi }_{a}\left( z\right) $ describes the Feshbach
coupling, and
\begin{equation}
\hat{V}_{ad}\left( z\right) =d_{ad}\hat{\Psi }^{\dag }_{a}\left(
 z\right) \hat{\Psi }^{\dag }_{a}\left( z\right) \hat{\Psi }_{d}\left
( z\right) , \hat{V}_{cd}\left( z\right) =d_{cd}\hat{\Psi }^{\dag
 }_{c}\left( z\right) \hat{\Psi }_{d}\left( z\right)  \label{Vadcd2B}
\end{equation}
are couplings of the open and closed channels, respectively, to the
product channels. The non-resonant interaction strength $U_{a}$,
 resonance
detuning $D_{c}$,  and the Feshbach coupling strength $g$ can be
 related to
parameters of atomic collisions and the waveguide \cite{Y05,Y06} [see
 Eq.\
(\ref{DimLess}) below].

Like in the two-channel case \cite{YBO06}, 2B problem can be described
by the 1D $T$ matrix
\begin{equation}
T_{1D}\left( k\right) =\left\lbrack U^{-1}_{\text{eff}}\left( k^{2}
/m\right) +{i\over 2}mk^{-1}\right\rbrack ^{-1} , \label{T1D}
\end{equation}
which depends on the relative momentum $k$ of two colliding atoms,
 where
the energy-dependent interaction strength $U_{\text{eff}}\left(
 E_{c}\right) $ incorporates effects of
all channels. The poles of $T_{1D}\left( k\right) $ on the positive
 imaginary axis, $k=i\kappa _{n}$,
correspond to 2B bound states (diatoms) with energies $-\kappa
 ^{2}_{n}/m$. They are
superpositions of the open, closed, and deactivation product
 channels. The
diatoms have finite size ($\sim \kappa ^{-1}_{n}$), although this
 model approximates the
closed-channel and the deactivation-product molecules to be
 infinitesimal
in size.

Substitution of the state vector for the three-atom system
\begin{eqnarray}
|\Psi _{3}\rangle =\biggl\{{1\over \sqrt{6}}\int d^{3}z\varphi
 _{0}\left( z_{1},z_{2},z_{3}\right) \hat{\Psi }^{\dag }_{a}\left(
 z_{1}\right) \hat{\Psi }^{\dag }_{a}\left( z_{2}\right) \hat{\Psi
 }^{\dag }_{a}\left( z_{3}\right)  \nonumber
\\
+\sum\limits^{}_{j=c,\{d\}}\int dz dz_{m}\varphi _{d}\left(
 z,z_{m}\right) \hat{\Psi }^{\dag }_{a}\left( z\right) \hat{\Psi
 }^{\dag }_{d}\left( z_{m}\right) \biggr\}|\text{vac}\rangle  ,
 \label{Psi3}
\end{eqnarray}
where $|$vac$\rangle $ is the vacuum state, into stationary
 Schr\"odinger
equation with the Hamiltonian (\ref{Hacd}) leads to coupled equations
for the wavefunctions of the open (three-atom) channel $\varphi
 _{0}\left( z_{1},z_{2},z_{3}\right) $,
the closed (atom-molecule) $\varphi _{c}\left( z,z_{m}\right) $, and
 the product channels
$\varphi _{d}\left( z,z_{m}\right) $. Elimination of the closed and
 product channels (like in
\cite{YBO06,Y06}) results in a single equation for $\varphi _{0}\left
( z_{1},z_{2},z_{3}\right) $, which
can be reduced to the Faddeev-Lovelace equation for the symmetric
transition amplitude $X\left( p,p_{0}\right) $,
\begin{eqnarray}
X\left( p,p_{0}\right) =2Z\left( p,p_{0}\right) +{m{ } ^{2}\over
 2\kappa { } ^{3}_{0}}\int dq Z\left( p,q\right) T_{1D}\left( k\left(
 q\right) \right) X\left( q,p_{0}\right)  \nonumber
\\
Z\left( p,q\right) ={2\kappa { } ^{3}_{0}\over \pi m}{1\over mE
+i0-p^{2}-pq-q{ } ^{2}} . \label{FadLov}
\end{eqnarray}
Here $k\left( q\right) =\sqrt{mE+i0-3q^{2}/4}$, $E$ is the total
 energy in the center-of-mass
system, $p_{n}=2\sqrt{\left( mE+\kappa ^{2}_{n}\right) /3}$  is  the
 relative momentum of the atom and diatom in
the state $n$, and  $n=0$ corresponds to the initial diatom state. The
probabilities of inelastic reflection and transmission with the diatom
transition to the state $n\neq 0$ can be expressed as
\begin{equation}
P_{\text{ref,tr}}\left( 0\rightarrow n\right) ={16\pi { } ^{2}\over
 9}{m^{2}W_{0}W{ } _{n}\over p_{0}p{ } _{n}} \left( {\kappa { }
 _{n}\over \kappa { } _{0}}\right) ^{3}|X\left( \mp p_{n},p_{0}\right
) |^{2} , \label{Preftr}
\end{equation}
where $\kappa _{0}<\kappa _{n}$  and $W_{n}$  is the contribution of
 the open channel into the
diatom state $n$ (see \cite{YBO06,Y06}). The deactivation rate
 coefficient is
\begin{equation}
K_{1D}={3p{ } _{0}\over 2m}\sum\limits^{}_{n\neq 0}\left\lbrack
 P_{\text{ref}}\left( 0\rightarrow n\right) +P_{\text{tr}}\left(
 0\rightarrow n\right) \right\rbrack  . \label{K1Dg}
\end{equation}
In the case of high deactivation energies,
\begin{equation}
|D_{d}|\gg \max(|d^{2}_{cd}/D_{c}|, |d_{ad}d_{cd}/g|, |d^{2}_{ad}
/U_{a}|, |E|, |D_{c}|, mU^{2}_{a}) ,
\end{equation}
$T_{1D}\left( k\right) $ has poles at $k=i\kappa _{d}\approx i\sqrt{m
|D_{d}|}$, corresponding to the deactivation
products. Other poles are approximately determined by the same cubic
equation as in the two-channel model \cite{YBO06,KD98},
\begin{equation}
\kappa ^{3}+{m\over 2}U_{a}\kappa ^{2}+mD_{c}\kappa +{1\over
 2}m^{2}D_{c}U_{a}-m^{2}|g|^{2}=0 . \label{kappa}
\end{equation}
Contributions of the product channels into the corresponding
weaker-bound diatoms ($\kappa _{n}\ll \kappa _{d}$, $n\neq d$) can be
 neglected. Whenever $U_{a}>0$ or
$D_{c}>2|g|^{2}/U_{a}$, when Eq.\ (\ref{kappa}) has a single real
 positive root,
the deactivation rate coefficient can be approximately expressed as
\begin{equation}
K_{1D}=\sum\limits^{}_{\{d\}}|\gamma _{ad}\varphi _{0}\left(
 0,0,0\right) +\gamma _{cd}\varphi _{c}\left( 0,0\right) |^{2}
 \label{K1D}
\end{equation}
in terms of the wavefunctions of the open and closed channels in the
origin
\begin{eqnarray}
\varphi _{0}\left( 0,0,0\right) ={\sqrt{3W_{0}\kappa { } _{0}}\over
 2\pi } \left\lbrack 1+{im{ } ^{2}\over 4\kappa { } ^{2}_{0}}\int dq
 {T_{1D}\left( k\left( q\right) \right) X\left( q,p_{0}\right) \over
 k\left( q\right) }\right\rbrack  \nonumber
\\
\varphi _{c}\left( 0,0\right) =-{mg\over \pi }\sqrt{{W_{0}\kappa { }
 _{0}\over 2}}\biggl\lbrack {1\over \kappa ^{2}_{0}+mD{ } _{c}}
 \nonumber
\\
+{m\over 2\kappa { } ^{2}_{0}}\int dq{T_{1D}\left( k\left( q\right)
 \right) X\left( q,p_{0}\right) \over U_{a}\left( k^{2}\left( q\right
) -mD_{c}\right) +2m|g|{ } ^{2}}\biggr\rbrack  . \nonumber
\end{eqnarray}
Here the transition amplitude $X\left( q,p_{0}\right) $ is a solution
 of  Eqs.\
(\ref{FadLov}) with $T_{1D}\left( k\right) $ (\ref{T1D}), where the
 energy-dependent
interaction strength is approximated by
\begin{equation}
U_{\text{eff}}\left( E_{c}\right) =U_{a}+{2|g|{ } ^{2}\over E_{c}
+i0-D{ } _{c}} , \label{Ueff}
\end{equation}
of the two-channel model \cite{YBO06}.

The coefficients in Eq.\ (\ref{K1D}) are expressed as
\begin{equation}
\gamma _{ad}=\left( {3^{5}4m\over |D_{d}|{ } ^{5}}\right) ^{1/4}
|d_{da}|U_{a}, \gamma _{cd}=2\left( {3^{3}m\over |D_{d}|{ }
 ^{5}}\right) ^{1/4}|d_{da}|g^{*} .
\end{equation}
The same deactivation rate can be obtained within another model,
assuming deactivation due to three-atom interactions,
\begin{eqnarray}
\hat{V}_{ad}\left( z\right) ={1\over 4\pi }\left( {|D_{d}|\over
 3m}\right) ^{1/4}\gamma _{ad}\hat{\Psi }^{\dag }_{a}\left( z\right)
 \hat{\Psi }^{\dag }_{a}\left( z\right) \hat{\Psi }^{\dag }_{a}\left(
 z\right) \hat{\Psi }_{a}\left( z\right) \hat{\Psi }_{d}\left(
 z\right)  \nonumber
\\
\hat{V}_{cd}\left( z\right) ={1\over 4\pi }\left( {3|D_{d}|\over
 4m}\right) ^{1/4}\gamma _{cd}\hat{\Psi }^{\dag }_{c}\left( z\right)
 \hat{\Psi }^{\dag }_{a}\left( z\right) \hat{\Psi }_{a}\left( z\right
) \hat{\Psi }_{d}\left( z\right)  , \label{Vadcd3B}
\end{eqnarray}
in place of two-atom ones (\ref{Vadcd2B}).

If the non-resonant case ($g=0$, $\varphi _{c}=0$) the problem is
 reduced to the LLMG
model  \cite{LL63,McGuire64}, which has an exact Bethe-ansatz
 solution. In
agreement with \cite{PSS}, $K_{1D}$  of Eq.\ (\ref{K1D}) is
 proportional to the
three-atom correlations (\ref{phiLLMG}), which does not describe
 suppression
of deactivation.

A physical sense of this effect can be explained by a simple 2B
analogy. Consider a collision of 1D atom and molecule with
 coordinates $y$ and
$x$, respectively. The wavefunction of this system has the form of
$\varphi \left( y,x\right) =\exp\left( ip_{0}\left( y-x\right) \right
) +R \exp\left( ip_{0}|y-x|\right) $, where $R$ is the reflection
 amplitude.
In the LLMG model atom-diatom reflection is forbidden, $R=0$, and the
 2B
analogy leads to $\varphi \left( 0,0\right) =1+R=1$. The 3B model
 results in Eq.\ (\ref{phiLLMG}),
which has a non-zero value too since the three atoms can approach each
other.

However, reflection becomes allowed when integrability of the LLMG
model is lifted e.g. by a Feshbach resonance \cite{YBO06}. Moreover,
reflection becomes the dominant channel at low collision energy,
 preventing
approaching of the atoms and leading to suppression of deactivation.

\begin{figure}
\includegraphics[width=3.375in]{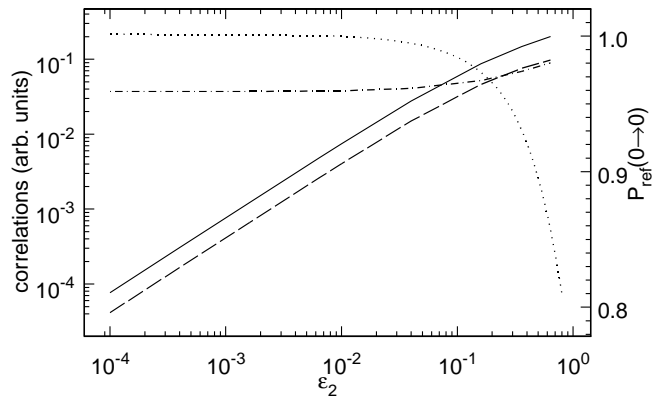}

\caption{Three-atom correlations for the open $|\varphi _{0}\left(
 0,0,0\right) |^{2}$  (dashed
line) and closed $|\varphi _{c}\left( 0,0\right) |^{2}$  (solid line)
 channels, as well as the elastic
reflection probability $P_{\text{ref}}\left( 0\rightarrow 0\right) $
(dotted line), calculated as functions of
the scaled collision energy $\epsilon _{2}$  for $U_{a}=0$ and
 $D_{c}=0$. The dot-dashed line
displays $|\varphi _{0}\left( 0,0,0\right) |^{2}$  within the LLMG
 model.} \label{Fig_E}

\end{figure}

\begin{figure}
\includegraphics[width=3.375in]{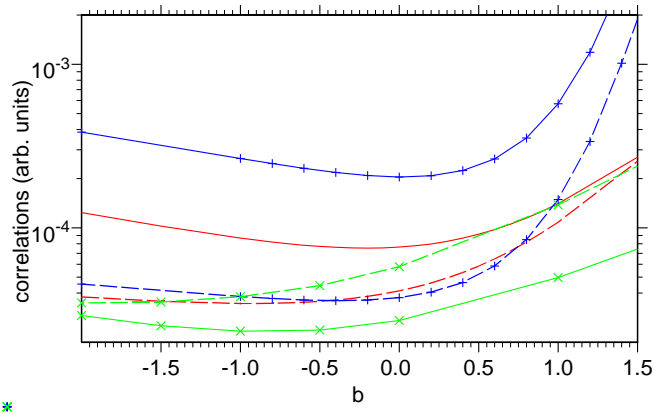}

\caption{Three-atom correlations for the open $|\varphi _{0}\left(
 0,0,0\right) |^{2}$  (dashed
lines) and closed $|\varphi _{c}\left( 0,0\right) |^{2}$  (solid
 lines) channels, calculated as
functions of the scaled detuning for $\epsilon _{2}=1\times 10^{-4}$
 and $u=1$ (pluses), $u=-1$
(crosses), and $u=0$ (no symbols).} \label{Fig_D}

\end{figure}

This hypothesis is confirmed by numerical calculations. As the
coefficients $\gamma $ are independent of the collision energy and
 resonance
detuning in both models, (\ref{Vadcd2B}) and (\ref{Vadcd3B}), the
 behavior
of the deactivation rate (\ref{K1D}) is determined by the three-atom
correlations for the open $|\varphi _{0}\left( 0,0,0\right) |^{2}$
 and closed $|\varphi _{c}\left( 0,0\right) |^{2}$  channels, see
Fig.\ \ref{Fig_E}. The results are expressed in terms of dimensionless
parameters: the non-resonant interaction strength $u$, the collision
 energy
$\epsilon _{2}=3p^{2}_{0}/\left( 4mD_{0}\right) $, and the detuning
 $b$, where $D_{0}$  is the energy scale.  In the
quasi-1D regime, when $p^{2}_{0}/m\ll \omega _{\perp }$  and the 3D
 elastic scattering length $a_{3D}$
does no exceed the transverse waveguide length $a_{\perp
 }=\left\lbrack 2/\left( m\omega _{\perp }\right) \right\rbrack ^{1
/2}$,  the
parameters can be expressed as \cite{Y06}
\begin{eqnarray}
D_{0}=m^{1/3}|g|^{4/3}=m^{1/3}\left( \omega _{\perp }a_{3D}\mu \Delta
 \right) ^{2/3}\beta ^{-4/3}_{1}\beta ^{2/3}_{2} \nonumber
\\
u=m^{1/3}|g|^{-2/3}U_{a}=2\left( m\omega ^{2}_{\perp
 }a^{2}_{3D}\right) ^{1/3}\left( \beta _{1}\beta _{2}\mu \Delta
 \right) ^{-1/3} \label{DimLess}
\\
b=D_{c}/D_{0}=\left\lbrack \mu \left( B-B_{0}\right) -\omega _{\perp
 }+Ca_{3D}\mu \Delta /\left( \beta _{1}a_{\perp }\right)
 \right\rbrack /D_{0} , \nonumber
\end{eqnarray}
with $\beta _{1}=1-Ca_{3D}/a_{\perp }$  and $\beta _{2}=1+C^\prime
 a_{3D}\mu \left( B-B_{0}-\Delta -\omega _{\perp }/\mu \right) ^{2}
/\left( 2a_{\perp }\omega _{\perp }\Delta \right) $. Here $\Delta $ is
the phenomenological resonance strength, $\mu $ is the difference
 between the
magnetic momenta of an atomic pair in the open and closed channels,
 $B-B_{0}$
is the detuning of the external magnetic field $B$ from its resonant
 value
$B_{0}$, $C\approx 1.4603$ \cite{O98}, and $C^\prime \approx 1.3062$
 \cite{Y06}. For example, in a
waveguide with the transverse frequency $\omega _{\perp }=50\times
 2\pi $  KHz, collisions remain
quasi-1D for the collision energy less then $2.4 \mu$K. In vicinity of
$b=0$, Eq.\ (\ref{DimLess}) gives $D_{0}=3 \mu $K, $u=0.024$, and $dB
/db=34$  mG for
the Na resonance at 907 G and $D_{0}=2.5 \mu $K, $u=0.12$, and $dB
/db=27$  mG for the
$^{87}$Rb resonance at 1007 G. Figure \ref{Fig_E} demonstrates that
 the
correlations and, therefore, the deactivation rate $K_{1D}$  decrease
proportionally to the collision energy for slow collisions, when the
 total
elastic reflection is approached. It is surprising that
 non-integrability
leads to the same low-energy behavior of deactivation rate of broad
molecules as in collisions of structureless particles
\cite{Gangardt_03,YB07}. At rather high collision energy, when the
 elastic
reflection probability decreases, correlations in the open channel
 follow
to the LLMG model. Deactivation suppression persists in a wide range
 of
the resonance detunings  and the non-resonant interaction strengths
(see
Fig.\ \ref{Fig_D}).

\begin{figure}
\includegraphics*[width=3.375in]{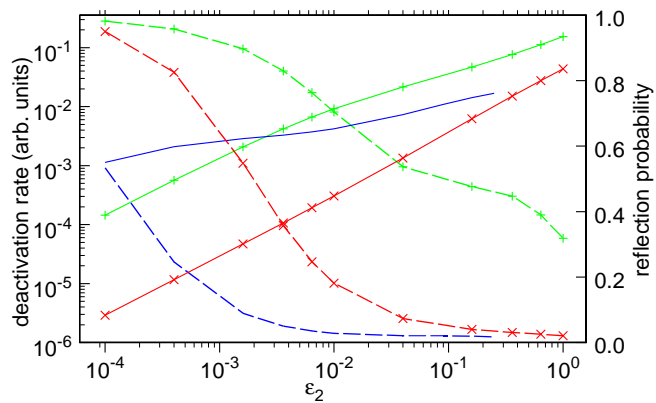}

\caption{Deactivation rate coefficient $K_{1D}$  (solid lines) and
 elastic
reflection probability $P_{\text{ref}}\left( 0\rightarrow 0\right) $
(dashed lines), calculated as functions
of the scaled collision energy $\epsilon _{2}$  for $b=0.4$, $u=10$
(crosses), $b=-2$, $u=-2$
(pluses), and $b=-4$, $u=-1$ (no symbols), within the two-channel
 model.}
\label{Fig_E2}

\end{figure}

The foregoing results are related to high deactivation energies
$|D_{d}|$. The case of low deactivation energies can be considered
 within the
two-channel model \cite{YBO06}. A Feshbach molecule can have two bound
states at $U_{a}<0$ and $D_{c}<2|g|^{2}/U_{a}$, when Eq.\
(\ref{kappa}) has two real
positive solutions, $\kappa _{1}>\kappa _{0}$. Collision with third
 atom can lead to
transitions between the corresponding states, and foreign deactivation
product states are not more necessary. This case is exactly described
 by
Eqs.\ (\ref{T1D}), (\ref{FadLov}), (\ref{Preftr}) and (\ref{Ueff}).
 The
deactivation rate is given by Eq.\ (\ref{K1Dg}), which includes now a
singe term ($n=1$) only. It is again proportional to the collision
 energy
for slow collisions (see Fig.\ \ref{Fig_E2}). Deactivation suppression
correlates with the increase in the elastic reflection probability in
this model too.

In summary, the integrable LLMG model does not predict suppression
of deactivation of broad 1D molecules, demonstrating their difference
from compact molecules. The deactivation becomes suppressed when the
symmetry of LLMG model is lifted by Feshbach resonance. This effect is
predicted by the two-channel model, as well as by two multichannel
models with different interactions, (\ref{Vadcd2B}) and
(\ref{Vadcd3B}).
Dibosonic Feshbach molecules in atom waveguides become thus relatively
stable, like difermionic ones in free space. Thus, both the presence
\cite{YBO06} and suppression of certain processes are among the
observable effect of non-integrability.

The author is very grateful to Yehuda Band for stimulating
discussion.

\end{document}